\documentclass[twocolumn,showpacs,preprintnumbers,amsmath,amssymb]{revtex4}
\usepackage{graphicx}% Include figure files
\usepackage{dcolumn}% Align table columns on decimal point
\usepackage{bm}% bold math
\usepackage{color}
\usepackage{amssymb}

\begin{document}

%\preprint{APS/123-QED}

\title{Holography from quantum cosmology}

 \author{ M. Rashki}
 \author{S. Jalalzadeh}
 \email{s-jalalzadeh@sbu.ac.ir}
\affiliation{Department of Physics, Shahid Beheshti University, G. C., Evin, Tehran, 19839, Iran}
   \date{\today}
%\maketitle
%-Title
%%%%%%%%%%%%%%%%%%%%%%%%%%%%%%%%%%%%%%%%%%%%%%%%%%%%%%%%%%%%%%%%%%%%%%%%%%%%%%%%
%+Abstract
\begin{abstract}
The Weyl-Wigner-Groenewold-Moyal formalism of deformation quantization
is applied to the closed  Friedmann-Lema\^itre-Robertson-Walker (FLRW) cosmological model. We show that the phase space average for the surface of the apparent
horizon is quantized  in units of the Planck's surface, and that the total entropy of the universe is also quantized. Taking into account these two concepts, it is
shown\ that   't Hooft conjecture on the cosmological holographic principle
(CHP) in radiation and dust dominated quantum universes is satisfied as a
manifestation of quantization. This suggests that the entire universe (not
only inside the apparent horizon) can be seen as a two-dimensional information structure encoded on the apparent horizon.

\end{abstract}
\maketitle
%-Abstract
%%%%%%%%%%%%%%%%%%%%%%%%%%%%%%%%%%%%%%%%%%%%%%%%%%%%%%%%%%%%%%%%%%%%%%%%%%%%%%
\section{Introduction}
  Deformation quantization, which is presented as Weyl-Wigner-Groenewold-Moyal phase space quantization, describes a quantum system in terms of the $c$-number (classical number) variables \cite{Moyal,WWGM}. Operators are mapped into the $c$-number functions
so that  their compositions could be obtained by the star product which is noncommutative but associative. Therefore, the observables  would be classical functions of the phase space. Quantum structure is constructed by replacing  point-wise products of classical observables
of the phase space, by star-product \cite{tillman,kont}.   The  product  of two smooth functions, say $f$ and $g$, on a Poisson manifold  is given by
 %%%%%%%%%%%%%%%%%%%
\begin{eqnarray}\label{star}
f\ast g:=\sum_{n=0}^\infty(i\hbar)^{n}\mathcal{C}_n(f,g)=fg+i\hbar\mathcal{C}_{1}(f,g)+{\mathcal
O}(\hbar^{2}),
\end{eqnarray}
%%%%%%%%%%%%%%%%%%%%
 where $\hbar$ plays the role of the deformation parameter. The first term denotes the common product of ${f}$ and ${g}$. Also, the coefficients $\mathcal{C}_n (f,g)$  are bi-differential operators, where their product is noncommutative
\cite{hirshfeld}. These coefficients satisfy the following properties
%%%%%%%%%%%%%%%%%%%%
\begin{eqnarray}\label{3 con.}
\begin{cases}
{\mathcal C}_{0}(f,g)=fg,\\
\mathcal{C}_{1}(f,g)-\mathcal{C}_{1}(g,f)=\{f,g\},\\
\sum_{i+j=n}\mathcal{C}_{i}(\mathcal{C}_{j}(f,g),h)=\sum_{i+j=n}\mathcal{C}_{i}(f,(\mathcal{C}_{j}(g,h)),
\end{cases}
\end{eqnarray}
%%%%%%%%%%%%%%%%%%%%
where $\{f,g\}$ denotes the Poisson bracket. In Eq. (\ref{3 con.}) the first
expression means that in the limit, $\hbar\rightarrow0$, the star product of $f$ and $g$ agrees with the point-wise products of these two functions. The second  expression shows that  at the lowest order of the deformation parameter,  the commutator  $[f,g]_{\ast}:=f*g-g*f$ tends to the Poisson bracket: $\lim _{\hbar\rightarrow0}\frac{1}{i\hbar}[f,g]_{*}=\{f,g\}$.
The last expression implies that, the star product  is associative: $(f*g)*h=f*(g*h)$.

One of the most important components of deformation quantization is the Wigner quasi-probability distribution  function (WF) \cite{WF,zachos1}. In fact, it is a generating function for all spatial autocorrelation functions of a given quantum mechanical
wave function \cite{Wig,fed}.
The WF in a $(2D)$-dimensional
phase space is given by
%%%%%%%%%%%%%%%%%%%%
\begin{eqnarray}\label{wigner f.}
\begin{array}{cc}
W_n(x,p):=\\
\frac{1}{(2\pi)^D }\int\psi_n^{\ast}(x-\frac{\hbar}{2}y)e^{-ip.y}\psi_n(x+\frac{\hbar}{2}y)d^Dy,
\end{array}
\end{eqnarray}
%%%%%%%%%%%%%%%%%%%%
 where $\psi_n$ is the state of the system. The distribution is real and the normalization is  expressed as $\int d^Dxd^Dp W_n(x,p)=1$. 
 %%%%%%%%%%%%%%%%%%%%
 
In flat  spaces, the special star product has long been
known. In this case, the components of the Poisson tensor $J^{ij}$ can
be considered constant. The coefficient $\mathcal{C}_{2}$ could be chosen
as antisymmetric so that

\begin{eqnarray}\label{c21}
\mathcal{C}_{2}(f,g)=\frac{1}{2}J^{ij}\partial_{i}f\partial_{j}g=\frac{1}{2}\{f,g\}.
\end{eqnarray}
In canonical coordinates, the poisson tensor $J$ is represented by the matrix
\begin{eqnarray}\label{alpha}
J=\begin{pmatrix}0
 & -I_{D} \\I_{D} & 0 \\
\end{pmatrix},
\end{eqnarray}
where $I_{D}$ is the $D\times D$ identity matrix.
 The higher order coefficients may be obtained by
exponentiation of $\mathcal{C}_{2}$. This procedure  yields the following Moyal star
product \cite{Moyal}
%%%%%%%%%%%%%%%%%%%%
\begin{eqnarray}\label{Moyal product}
\begin{array}{cc}
f(x,p)\ast_{\textsc{m}}g(x,p):=f\exp\left(\frac{i\hbar}{2}(\overleftarrow{\partial_{q}}\overrightarrow{\partial_{p}}-\overleftarrow{\partial_{p}}\overrightarrow{\partial_{q}})\right)g\\
=f(x+\frac{i\hbar}{2}\vec\partial_p,p-\frac{i\hbar}{2}\vec\partial_x)g(x,p),
\end{array}
 \end{eqnarray}
%%%%%%%%%%%%%%%%%%%%
where in the last step we used the Bopp shift argument. 
An alternative integral representation of the Moyal star product is given by \cite{Arratia}
\begin{eqnarray}\label{product1}
\begin{array}{cc}
f*_\textsc{m}g=\\
\frac{1}{(2\pi\hbar)^{4D}}\int d^{2D} vd^{2D} v'f( v)g(v')e^{\frac{i}{\hbar}(
u^tJ v+ v^tJ v'+ v'^tJ u)},
\end{array}
\end{eqnarray}
where $u=( x, p)^t$, $ v=( x', p')^t$ and
$ v'=( x'', p'')^t$. As a direct consequence, the Moyal star product
is a non local product. As a result, we have
\begin{eqnarray}\label{product2}
\int  d^Dxd^Dpf*g=\int  d^Dxd^Dpg*f=\int  d^Dxd^Dpfg.
\end{eqnarray}
The WF is closely tied to the wave function.
 Therefore,  it is necessary to define the phase space integrals corresponding to the expectation values of the operator formalism.  The expectation value or ``phase space average'' of phase space function, say
 $A(x,p)$, is given by
 %%%%%%%%%%%%%%%%%%%%
 \begin{eqnarray}\label{expect. obser.}
 \begin{array}{cc}
\left \langle A(x,p)\right \rangle= 
  \int W(x,p)\ast A(x,p)d^Dxd^Dp=\\
  \int A(x,p)\ast\ W(x,p)d^Dxd^Dp,
 \end{array}
 \end{eqnarray}
 where in the last step we have used the property expressed by Eq. (\ref{product2}).
The $ \ast_{\textsc{m}}$-genvalue equation for WF is given by \cite{Wig}
%%%%%%%%%%%%%%%%%%%%
\begin{eqnarray}\label{star.gen value}
{H}\ast_{\textsc{m}}W_n(x,p)={E_n}W_n(x,p),
\end{eqnarray}
or equivalently
\begin{eqnarray}\label{S1}
H(x+\frac{i\hbar}{2}\vec\partial_p,p-\frac{i\hbar}{2}\vec\partial_x)W_n(x,p)=E_nW_n(x,p),
\end{eqnarray}
%%%%%%%%%%%%%%%%%%%%
where ${H}$ is the Weyl correspondence to the Hamiltonian and ${E_n}$ is the spectrum of energy.
The dynamical equations in  this picture are given by Moyal's equation 
\begin{eqnarray}\label{Li}
\frac{\partial f}{\partial t}=\frac{1}{i\hbar}[H,f]_{*_{\textsc m}}.
\end{eqnarray}
In
fact, it is the generalization of Liouville's theorem of classical mechanics.
The Moyal dynamical equation is similar to the Heisenberg's equation of
motion for operators. But here, $H$ and $f$, as was said previously, are phase
space functions, not operators. Another point in this formulation of quantum
mechanics is  the absence of the wave function.
  This    plays an important role in the construction of quantum cosmology. In quantum cosmology, problems occur in two ways. Firstly when the Copenhagen
interpretation is implemented, and secondly when the  working tool is the
wave function.  In the former, the observer itself is also an element of the quantum cosmology, where the Copenhagen interpretation requires an external observer, while the whole universe has  nothing external to it.
For the latter, we must ask, how  is it possible to construct a wave packet 
which would peak around the classical trajectories in the configuration space; the wave function
describing this universe must approach a wave packet that characterizes  the
presently observed cosmological data. The  advantage of  deformation
quantization  is that it makes quantum cosmology look like the
Hamiltonian formalism of cosmology. This is done by avoiding the operator formalism. 
 %%%%%%%%%%%%%%%%%%%%%%%%%%%%%%%%%%%%%%%%

 The  holographic principle  is a feature of string theory and in principle implies that the degrees of freedom in a spatial region can all be encoded on its boundary. Note that, the holographic principle was first proposed by Gerard 't Hooft \cite{'t Hooft1},  where it is worth seeing   \cite{suskind 1} if interested in a string theory  interpretation.
The holographic principle has since been applied in the context of pre big-bang scenarios \cite{pre. big bang}, singularity problem \cite{singular prob.}, and inflation \cite{inflation}, typically for a flat universe.
Also, it is investigated regarding the standard big-bang cosmology by Fischler and Susskind (FS)   \cite{suskind 2}.  They have found that  if our universe   is flat or open, it obeys this principle.  This (FS) version of CHP demands that the  entropy contained in a volume of particle horizon should not  exceed the area of the horizon in Planck units. 
Lately, there have been two further  proposals for the completion of the holographic principle by Easther and Lowe, based on the second law of thermodynamics
\cite{EL holography},  and by Bak and Rey, using the cosmological apparent horizon instead of the particle horizon \cite{Bak ray 1}. In both of these completions, the closed universe also obeys the holographic principle naturally. Therefore, these proposals are perhaps  more natural compared to  the FS proposal.

In this paper we investigate the quantum cosmology of a closed FLRW universe, filled with  radiation  or  dust. In the first step, we investigate
the deformation quantization of the model. Using WF we show that the deformed cosmology
predicts a good agreement with the corresponding classical cosmology. Also,
we demonstrate that the phase space average of apparent horizon is quantized.
This leads us to conclude that the total entropy
of radiation or a dust dominated quantum universe satisfies 't Hooft conjecture. The paper consists of the following sections. In Section II we present the classical model. Section III provides quantum cosmological description of the model and quantization rules. In Section IV, we summarize our results.

 %%%%%%%%%%%%%%%%%%%%%%%%%%%%%%%%%%%%%%%%%%%%%%%%%%%%%%%%%%%%%%%%%%%%%%%%%%%%%%
 
\section{The Classical Model}

A useful cosmological model that agrees well with observations   is the homogeneous and isotropic FLRW universe. In this model the line element  {}  for a closed universe   is given
by
%%%%%%%%%%%%%%%%%%%%%%%%%%%%%%%%%%%%%%%%%%%%%%%%%%
 \begin{eqnarray}\label{line frw}
ds^2=-N^2(t)dt^{2}  
+a^{2}(t)d\Omega_{(3)} ^{2},
\end{eqnarray}
%%%%%%%%%%%%%%%%%%%%%%%%%%%%%%%%%%%%%%%%%%%%%%%%%%% 
where $N(t)$ is the lapse function, $a(t)$ is the scale factor and  $ d\Omega_{ (3)} ^{2}$ is the standard line element of the unit three-sphere. The  action functional which consists of a gravitational part and a matter part when the matter field is considered as a perfect fluid,  is given by \cite{haw-ellis}

%%%%%%%%%%%%%%%%%%%%%%%%%%%%%%%%%%%%%%%%%%%%%%%%%
  \begin {eqnarray}\label{action}
  \begin{array}{cc}
  {\mathcal I}=\frac{M^{2}_{\textsc{p}}}{2}\int_{\mathcal{M}} \sqrt{-g}Rd^{4}x  +M^{2}_{\textsc{p}}\int _{\partial \mathcal{M}}\sqrt{g^{(3)}}Kd^{3}x\\ -\int _{\mathcal{M}}\sqrt{-g}\rho d^{4}x, 
\end{array}
\end{eqnarray}
%%%%%%%%%%%%%%%%%%%%%%%%%%%%%%%%%%%%%%%%%%%%%%%%%%%
where $ M_{\textsc{p}}=\frac{1}{\sqrt{8\pi G}}=\frac{1}{L_{\textsc{p}}}$ is the reduced Planck's mass  in
natural units $(c=\hbar=k_\textsc{b}=1)$, 
$\mathcal{M}=I\times S^{3}$ is the spacetime manifold, $ \partial\mathcal{M}$
is equal to $S^{3}$,  $K$ is the trace of extrinsic curvature  of the spacetime boundary and the overdot  denotes differentiation with respect to $t$. 
If we assume a universe filled with  non-interacting dust $\rho=\rho_{0m}(a/a_0)^{-3}$ and radiation $\rho=\rho_{0\gamma}(a/a_0)^{-4}$,
and  redefining the scale factor and the lapse function as
%%%%%%%%%%%%%%%%%%%%
\begin{eqnarray}\label{a,n}
\begin{cases}
 N(t)=12\pi^{2}M_{\textsc{p}}a(t)\tilde N(t),\\
a(t)=x(t)+\frac{M}{12\pi^{2}M^{2}_{\textsc {p}}}=x-x_{0},
\end{cases}
\end{eqnarray} 
%%%%%%%%%%%%%%%%%%%%%%%%%%%%%%%%%%%%%%
the total Lagrangian will be \cite{{jalalzadeh-moniz}}
%%%%%%%%%%%%%%%%%%%%
\begin{eqnarray}\label{lag}
\mathcal{L}=-\frac{1}{2\tilde N}M_{\textsc{p}}\dot x^2
+\frac{\tilde N}{2}M_{\textsc{p}}\omega^{2}x^2-\mathcal{E}\tilde N,
\end{eqnarray}
%%%%%%%%%%%%%%%%%%%%%%%%%%%%%   
where we have defined   
%%%%%%%%%%%%%%%%%%%%%%%%%%%%%%%%%%%%%%%%%%%%%%%%%%%%%%%
\begin{eqnarray}\label{e& omega}
\mathcal{E}=12\pi^{2}\mathcal{N}_{\gamma} M_{\textsc{p}}+ \frac{M^{2}}{2M_{\textsc{p}}},  \hspace{.2cm}
\omega=12\pi^{2}M_{\textsc{p}}.
\end{eqnarray}
%%%%%%%%%%%%%%%%%%%%%%%%%%%%%%%%%%%%%%%%%%%%%%%%%%%%%%%%%%
Besides, we introduce $M$ and $\mathcal{N}_{\gamma}$ as 
%%%%%%%%%%%%%%%%%%%%
\begin{eqnarray}\label{mass, number }
\begin{cases}
M=\int_{\partial M}\sqrt{g^{(3)}}\rho_{0m}a^{3}_{0}d^{3}x=2\pi^2 \rho_{0m}a_0^3,  \\
\mathcal{N}_{\gamma}=\int_{\partial M}\sqrt{g^{(3)}}\rho_{0\gamma}a^{4}_{0}d^{3}x=2\pi^2\rho_{0\gamma}a_0^4,
\end{cases}
\end{eqnarray}
%%%%%%%%%%%%%%%%%%%%%%
 where, $M$ is the total mass of the dust content of the universe and $\mathcal{N}_{\gamma}$  could be related to the total entropy of radiation, see Eq.  (\ref{radi&entropy}).  
 The conjugate momentum  to the shifted  scale factor $x$ and the primary constraint  are given by
%%%%%%%%%%%%%%%%%%%%%%%%%%
\begin{eqnarray}\label{primary constraint}
\begin{cases}
\Pi_{x}=\frac{\partial\mathcal{L}}{\partial\dot x}=-\frac{M_{\textsc{p}}}{\tilde N}\dot x,  \\
\Pi_{\tilde N}= \frac{\partial\mathcal{L}} {\partial \dot {\widetilde{N}} }=0.
\end{cases}
\end{eqnarray}
%%%%%%%%%%%%%%%%%%%%%%%
 Consequently,  the Hamiltonian  corresponding to
Lagrangian (\ref{lag}) will be
%%%%%%%%%%%%%%%%%%%%%%%%
\begin{eqnarray}\label{canonical.hamilton}
H\mathcal:=-{\tilde N\mathcal{H}}=-\tilde N\left[ \frac{1}{2M_{\textsc{p}}}\Pi^2_{x}\ +\frac{1}{2}M_{\textsc{p}}\omega^{2} x^{2}-\mathcal{E}\right].
\end{eqnarray}
%%%%%%%%%%%%%%%%%%%%%%%%
In  Hamiltonian (\ref{canonical.hamilton}), $\tilde N$ is a Lagrange multiplier, therefore  it enforces the Hamiltonian constraint
%%%%%%%%%%%%%%%%%%%%%%%%
\begin{eqnarray}\label{hamilton const.}
\mathcal{H}= \frac{1}{2M_{\textsc{p}}}\Pi^2_{x}\ +\frac{1}{2}M_{\textsc{p}}\omega^{2} x^{2}-\mathcal{E=}0.
\end{eqnarray}
Eq. (\ref{hamilton const.}) for any value of $\mathcal{E}$ shows
the elliptical patterns in 2-dimensional phase space. By choosing the gauge $\widetilde{N}=\frac{1}{\omega}$
the Hamiltonian equations of motion will be
\begin{eqnarray}\label{PE}
\dot x=\{x,H\}=\frac{1}{\omega M_{\textsc{p}}}\Pi_x,\hspace{.1cm}\dot\Pi_x=\{\Pi_x,H\}=-\omega
M_{\textsc{p}}x,
\end{eqnarray}
which leads us to 
\begin{eqnarray}\label{EQ}
\begin{cases}
x(t)=x(t_0)\cos(t)+\frac{1}{\omega M_{\textsc{p}}}\Pi_x(t_0)\sin(t),\\
\Pi(t)=\Pi(t_0)\cos(t)-\omega M_{\textsc{p}}x(t_0)\sin(t).
\end{cases}
\end{eqnarray}
If we assume that the origin of cosmic time is $t_0=0$ and $x(0)=x_0$, where $x_0$
is defined in (\ref{a,n}), we obtain the well-known  classical solution  
\begin{eqnarray}
\begin{cases}
a(t)=\frac{a_{\text{max}}}{1+\sec\phi}[1-\sec\phi \cos(t+\phi)],\\
a_{\text{max}}:= \frac{M}{12\pi^{2}M^{2}_{\textsc{p}}}+(\frac{2\mathcal{E}}{M_{\textsc{p}} \omega^{2}})^{\frac{1}{2}},\\
\cos(\phi):=\frac{M}{\sqrt{2 \mathcal{E}M_{\textsc{p}}}},
\end{cases}
\end{eqnarray}
%%%%%%%%%%%%%%%%%%%%%%%%%%%%%%%%%
where $a_{\text{max}}$ is the maximum radius of the closed universe.

\section{ Deformation Quantization  }
The deformation quantization of this simple model is accomplished
straightforwardly by replacing the ordinary products of the observables in phase
space by Moyal product. Therefore,  Hamiltonian constraint   (\ref{hamilton const.}) becomes the Moyal-Wheeler-DeWitt (MWDW) equation by replacing
the classical Hamiltonian (\ref{hamilton const.}) with its deformed counterpart  \cite{cordero}
%%%%%%%%%%%%%%%%%%%%
\begin{eqnarray}\label{MWDW}
\begin{array}{cc}
\mathcal{H}\ast_{\textsc{m}}W_{n}(x,\Pi_x)=\\\mathcal H\left(x+\frac{i\hbar}{2}\vec\partial_{\Pi_x},\Pi_x-\frac{i\hbar}{2}\vec\partial_x\right)W_n(x,\Pi_x)=0.
\end{array}
\end{eqnarray}
For simple Hamiltonian defined in (\ref{hamilton const.}), this equation has turned into two simple PDEs \cite{zachos1, hirshfeld}. The
imaginary part of this equation, restricts WF to depend on $\frac{1}{2}(\frac{1}{M_{\textsc{p}}}\Pi^2_{x}\ +M_{\textsc{p}}\omega^{2} x^{2})$. The real part  yields Laguerre's equation. Hence, one can easily find the following solution of WDWM equation
for the closed FLRW cosmology  
\begin{eqnarray}\label{wigner function}
\begin{array}{cc}
W_n(x,\Pi_{x})=\\\frac{(-1)^{n}}{\pi}\exp(-{{\frac{\Pi^2_{x}}{M_{\textsc{p}}\omega}-M_{\textsc{p}}\omega x^{2}}}) L_{n}\left(\frac{2\Pi^2_{x}}{\omega M_{\textsc{p}}}+2M_{\textsc{p}}\omega x^{2}\right),
\end{array}
\end{eqnarray}
  where $L_{n}(z)$ represents the Laguerre polynomials. Figure 1  shows the WF  of
the model for the third excited state. It will be observed that there exists
a  pattern
for the extrema in the vicinity of classical loci defined in Eq. (\ref{hamilton const.}).
Also, the Moyal evolution equations (\ref{Li}) will be
\begin{eqnarray}\label{ME}
\begin{cases}
\dot x=\frac{1}{i}(x*_{\textsc{m}}H-H*_{\textsc{m}}x)=\frac{1}{\omega M_{\textsc{p}}}\Pi_x,\\
\dot\Pi_x=\frac{1}{i}(\Pi_x*_\textsc{m}H-H*_\textsc{m}\Pi_x)=-\omega
M_{\textsc{p}}x.
\end{cases}
\end{eqnarray}
The solutions of the above deformed cosmology are
\begin{eqnarray}\label{EQ1}
\begin{cases}
x(t)=x(t_0)\cos(t)+\frac{1}{\omega M_{\textsc{p}}}\Pi_x(t_0)\sin(t),\\
\Pi(t)=\Pi(t_0)\cos(t)-\omega M_{\textsc{p}}x(t_0)\sin(t).
\end{cases}
\end{eqnarray}
These look similar to the classical equations of motion (\ref{EQ}). 
These equations of motions show that the functional form of WF is preserved
along classical phase space trajectories.

\begin{figure}[htb]
\begin{center}
\includegraphics[width=6cm]{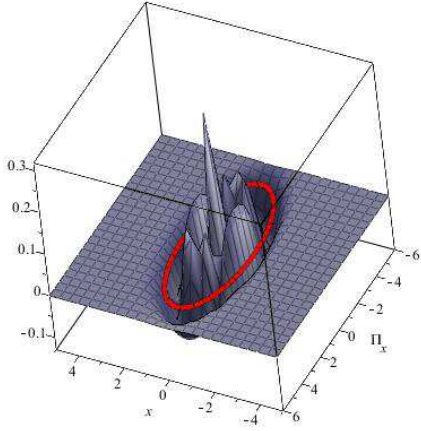}
%\hfill
%\centering
%\includegraphics[width=5cm]{wignercont.eps}
\caption{{ Wigner function for third excited
state, $n=3$, ($M_{p}=\frac{1}{2\pi}$). The corresponding classical trajectory is denoted by redline loci.}}
\end{center}
\end{figure}

%%%%%%%%%%%%%%%%%%%%
 Let us define in the unconstrained phase space, the complex-valued holomorphic  functions   
 %%%%%%%%%%%%%%%%%%%%%%%%%%%%%%%%%%%
 \begin{eqnarray}\label{holomo.}
\begin{cases}A=\sqrt{\frac{\omega M_{\textsc{p}}}{2}}(x+\frac{i\Pi_{x}}{\omega M_{\textsc{p}}}),\\
\bar A=\sqrt{\frac{\omega M_{\textsc{p}}}{2}}(x-\frac{i\Pi_{x}}{\omega M_{\textsc{p}}}).  \\ 
\end{cases}
\end{eqnarray}
%%%%%%%%%%%%%%%%%%%%%%%%%%%%%%%%%%%
 Then,     classical Hamiltonian  (\ref{hamilton const.}) will be%%%%%%%%%%%%%%%%%%%%%%%%%%%%%%%%%%
\begin{eqnarray}\label{hamiltonian holomorph}
\mathcal{H}=\omega  A  \bar A-\mathcal{E}.
\end{eqnarray}
%%%%%%%%%%%%%%%%%%%%%%%%%%%%%%%%%%% 
On the other hand, the Moyal commutation relation between these new variables is 
%%%%%%%%%%%%%%%%%%%%
\begin{eqnarray}\label{commiut.}
[A,\bar A]_{\ast_\textsc{m}}=A\ast_{\textsc{m}}\bar A -\bar A \ast_\textsc{m}A = 1,
\end{eqnarray}
where the  Moyal star product is redefined as  $ \ast_{\textsc{m}}:= e^{\frac{1}{2}\left(\overleftarrow{\partial_{A}} \overrightarrow{\partial_{\bar A}}-\overleftarrow{\partial_{\bar A}}\overrightarrow{\partial_{A}}\right)}$
\cite{hirshfeld}.  The Moyal star product between $A$ and $\bar A$ leads
us to the following relation between star and ordinary products of holomorphic
variables %%%%%%%%%%%%%%%%%%%%
\begin{eqnarray}\label{A bar A}
 \bar A \ast_{\textsc{m}}A= \bar A e^{\frac{1}{2}\left(\overleftarrow{\partial_{A}} \overrightarrow{\partial_{\bar A}}-\overleftarrow{\partial_{\bar A}}\overrightarrow{\partial_{A}}\right)}  A= \bar A A - \frac{1}{2}. 
\end{eqnarray}
%%%%%%%%%%%%%%%%%%%%
Consequently, by combining Eq. (\ref{A bar A})   and Eq. (\ref{hamiltonian holomorph}) we obtain the Hamiltonian for the model as
%%%%%%%%%%%%%%%%%%%%
\begin{eqnarray}\label{moyal.hamilton}
\mathcal{H}= \omega(\bar{A}*_\textsc{m}A+\frac{1}{2})-\mathcal{E.}
\end{eqnarray}
%%%%%%%%%%%%%%%%%%%%%%%%%%%%
In addition, the Wigner function (\ref{wigner function}), in terms of the holomorphic variables will be
\begin{eqnarray}\label{wigner. holomo.}
W_n(A,\bar A)=\frac{1}{n!}(\bar A)^{n} \ast_\textsc{m}W_0\ast_\textsc{m}(A)^{n},
\end{eqnarray}
%%%%%%%%%%%%%%%%%%%%
where $W_0=2e^{-2A\bar A}$ denotes the ground state
of the WF.  Note that for the ground state we have $A\ast_\textsc{m}{W}_0=0=W_0\ast_\textsc{m}\bar
A$.
Now, the MWDW Eq. (\ref{MWDW})  will be 
 %%%%%%%%%%%%%%%%%%%%%%%%%%%%%%%%%%%
\begin{eqnarray}\label{MWDW.solve}
\begin{array}{cc}
\mathcal{H}*_\textsc{m}W_n=\omega(\bar{A}*_\textsc{m}A+\frac{1}{2}-\mathcal{E})*_\textsc{m}W_n\\=   \left[\omega(n+\frac{1}{2})-\mathcal{E}\right]W_n=0,
  \end{array}
\end{eqnarray}
%%%%%%%%%%%%%%%%%%%%%%%%%%%%%%%%%%% 
 which leads  to 
%%%%%%%%%%%%%%%%%%%%%%%%%%%%%%%%%%%
\begin{eqnarray}\label{enrge. quant.}
\mathcal{E}_{n}=\omega(n+\frac{1}{2}).
\end{eqnarray}

%%%%%%%%%%%%%%%%%%%%%%%%%%

\subsection{Cosmological holographic principle in a radiation dominated universe}

  Let us first assume that the universe is  radiation dominated, where $M=0$. In
  this case, Eq. (\ref{enrge. quant.}) and  definition of $\mathcal
E$ in (\ref{e& omega}) give
\begin{eqnarray}\label{ss1}
{\mathcal N}_\gamma=n+\frac{1}{2}.
\end{eqnarray}
%%%%%%%%%%%%%%%%%%%%
As was mentioned at the beginning of this section, $\mathcal{N}_{\gamma}$ could be related to the total entropy of radiation. Recalling the relation of the energy density of radiation  $\rho_\gamma $, the entropy density ${s_\gamma}$ and the scale factor ${a}$ with  temperature,  $\rho_{\gamma}=\frac{\pi^{2}}{30}gT^{4}$,
$ s_{\gamma}=\frac{4}{3}\frac{\rho}{T}$, $a(t)\sim\frac{1}{T}$  \cite{mukhan},
and using these relations in
definition of ${\mathcal N}_\gamma$ in (\ref{mass, number }), we find
%%%%%%%%%%%%%%%%%%%%
\begin{eqnarray}\label{radi&entropy}
\mathcal{N}_{\gamma}=\left(\frac{5\times3^{5}}{2^{8}\pi^{4}g}\right)^{\frac{1}{3}}(S_{\gamma})^{\frac{4 }{3}},
\end{eqnarray}
%%%%%%%%%%%%%%%%%%%%
where $S_\gamma=2\pi^2a^3s_\gamma$ denotes the total entropy and  $g$ is the internal degrees of freedom. Now, by inserting  (\ref{ss1}) into Eq. (\ref{radi&entropy}),
we obtain
%%%%%%%%%%%%%%%%%%%%
\begin{eqnarray}\label{entropy quantize}
 S_{\gamma} =\left(\frac{2^8\pi^4g}{5\times3^5} \right)^\frac{1}{4} \left(n+\frac{1}{2}\right)^\frac{3}{4},
\end{eqnarray}
%%%%%%%%%%%%%%%%%%%%
which shows that the total entropy of radiation is quantized.  Let us now deal with the relation between the total entropy and the phase space average of the apparent horizon. First note that in definition (\ref{a,n}), for a
radiation dominated universe, we have  $x(t)=a(t)$. Hence Eq. (\ref{holomo.})
leads us to obtain the scale factor in terms of holomorphic variables $a(t)=\frac{1}{\sqrt{2M_\textsc{p}\omega}}(A+ \bar A)$.
One can easily show that the phase space average of biquadratic   scale factor is  
 %%%%%%%%%%%%%%%%%%%%
\begin{eqnarray}
\begin{array}{cc}
\left\langle a^{4} \right\rangle=
\frac{1}{4M_\textsc{p}^2\omega^2}\left\langle (A+\bar A)^4\right\rangle\\
 = \frac{3}{ 2M_\textsc{p}^2 \omega^2}\left(n^2+n+\frac{1}{2}\right).
 \end{array}
\end{eqnarray}
  On the other hand, the apparent horizon of a radiation dominated
 universe  is given by,  $R_{\text{ah}}^2=(\mathrm{H}^{2}+\frac{ 1}{a^{2}})^{-1}={\frac{6\pi^{2}M^{2}_\textsc{p}}{\mathcal{N}_{\gamma}}}a^{4}$,
where ${H}$ is the Hubble parameter. Therefore  the phase space average of the area for the apparent horizon becomes
%%%%%%%%%%%%%%
\begin{eqnarray}\label{sur.apparent}
  \langle\mathcal{A}_{\text{ah}}\rangle :=4\pi\langle R^2_{\text{ah}}\rangle=\frac{L^2_\textsc{p}}{4\pi}\left(1+\frac{n^2}{n+\frac{1}{2}}\right),
  \end{eqnarray}
  %%%%%%%%%%%%%%%%%%%%%%%%%%%%%%%%%%%%%%%%%%%%%%%%%%%%%%
where $L_\textsc{p}$ is the reduced Planck's length. Hence, the phase space
average of the apparent horizon is quantized.  By comparing Eqs. (\ref{entropy quantize}) and (\ref{sur.apparent}) for large values of the quantum number $n$
we obtain   
%%%%%%%%%%%%%%%%%%%%
\begin{eqnarray}\label{tHooft entropy}
S_{\gamma}\simeq g^\frac{1}{4}\left(\frac{\langle\mathcal{A}_{\text{ah}}\rangle}{4G}\right)^{\frac{3}{4}}.
\end{eqnarray}
%%%%%%%%%%%%%%%%%%%%
The above equation is in the form  conjectured
by 't Hooft \cite{'t Hooft1}.

%%%%%%%%%%%%%%%%%%%%%%%%%%%%%%%%%%%%%%%%%%%%%%%%%%%%%%%%%%%%%%%%%%%%%%    
\subsection{'t Hooft conjecture in a dust dominated universe }
Let us now return to a universe filled  only with dust, (${\mathcal{N}_{\gamma}}=0$). In this case, comparing Eqs. (\ref{e& omega}) and (\ref{enrge. quant.}) implies the following quantization
 rule for the total mass of the universe
%%%%%%%%%%%%%%%%%%%%
\begin{eqnarray}\label{mass.quantize}
 M= \sqrt{24 \pi^{2}(n+\frac{1}{2})}M_{\textsc p}.
\end{eqnarray}
  %%%%%%%%%%%%%%%%%%%%
We now estimate the total entropy of the dust dominated universe. Consider the case where a system has a total of $\Omega$ states of equal likelihood. Then the entropy will be
 \begin{eqnarray}\label{entropy n part.}
 S=\ln(\Omega), \hspace{.3cm}(k_\text{B}=1).
 \end{eqnarray}
  Further, let us assume that all the particles are identical. Then $\Omega=\tilde
  n^{N}$,
 where $\tilde n$ is the number of states accessible to a single particle,  hence 
 \begin{eqnarray}\label{entropy omega}
 S=N\ln(\tilde n).
\end{eqnarray}  
 Evaluating the one particle phase space, one finds \cite{kittel}  for an ideal  gas with $N$  free particles       
  %%%%%%%%%%%%%%%%%%%%
\begin{eqnarray}\label{entropy ideal gas}
S_{(\text{ideal})}=N\ln\left(\frac{V}{N}(\frac{\text{m}T}{2\pi})^{\frac{3}{2}}e^{\frac{5}{2}}\right),
\end{eqnarray}
where  $V$ is the volume and $m$ denotes the mass of particles.
 %%%%%%%%%%%%%%%%%%%%
    For the case of
a continuous fluid, let us rewrite Eq. (\ref{entropy ideal gas}). To this end, we
consider an ideal gas contained within a small volume
element $dV$. The number of particles inside $dV$ is
%%%%%%%%%%%%%%%%%%%%
\begin{eqnarray}\label{dN}
dN=\frac{\rho}{\text{m}}dV.
\end{eqnarray} 
  %%%%%%%%%%%%%%%%%%%%
 Inserting  expression (\ref{dN}) into Eq. (\ref{entropy ideal gas}), the entropy associated with the volume element, in terms of the density
of the fluid, can be written as 
 %%%%%%%%%%%%%%%%%%%%
\begin{eqnarray}
dS^{(\text{dust})}=\frac{\rho}{\text{m}}\ln\left(\frac{KT^{\frac{3}{2}}}{\rho}\right)dV,
\end{eqnarray}
%%%%%%%%%%%%%%%%%%%%
where $K=(\frac{m^{5}e^{5}}{2\pi})^{\frac{1}{2}}$ \cite{gron}. For a dust dominated universe, the density and temperature are   $\rho=\rho_{0}(\frac{a}{a_{0}})^{-3}$ and $T=T_{0}(\frac{a}{a_{0}})^{-2}$. Hence we have $S^{(\text{dust})}=\ln(KT^{\frac{3}{2}}_{0}/\rho_{0})N$.
 We use the simple approximation
%%%%%%%%%%%%%%%%%%%%
\begin{eqnarray}\label{entropy dust}
S^{(\text{dust})}\simeq N=\frac{M}{\text{m}},
\end{eqnarray} 
which is accurate  within two
orders of magnitude because, as noted
by Fermi, all large logs are less than a thousand even in
cosmology. 
%%%%%%%%%%%%%%%%%%%%
Therefore from Eqs. (\ref{mass.quantize}) and (\ref{entropy dust})  we obtain
%%%%%%%%%%%%%%%%%%%%
\begin{eqnarray}\label{ent.dust.quant}
S_{n}^{(\text{dust})}\simeq  \frac {M_{\textsc P}}{\text{m}}\pi \sqrt{24(n+\frac{1}{2})}.
 \end{eqnarray}
 %%%%%%%%%%%%%%%%%%%%
 Let us investigate 't Hooft conjecture for this
model. The apparent horizon of a dust dominated universe using the definition of total mass in
Eq. (\ref{mass, number }) and the Friedmann equation  is given by %%%%%%%%%%%%%%%%%%%%%%%%%%%%%%%%%%%%%%%%%%%%%%%%%%%
\begin{eqnarray}\label{A.H}
R^2_{\text{ah}}=\frac{6\pi^{2}M_{\textsc P}^2}{M}a^{3}.  
\end{eqnarray}
%%%%%%%%%%%%%%%%%%%%%%%%%%%%%%%%%%%%%%%%%%%%%%%%%%%%%%%%%%%%%%%%% 
 Moreover, using definition  (10), the phase space average of the cubic   scale
 factor   will be
 %%%%%%%%%%%%%%%%%%%%
 \begin{eqnarray}\label{exp. value. qub.scal}
 \begin{array}{cc}
  \left \langle a^{3}\right \rangle=\left\langle (x-x_{0})^{3}\right\rangle=\\ \left\langle x^{3} \right\rangle -3x_{_{0}} \left\langle x^{2} \right\rangle+3x^{2}_{0}\left\langle x \right\rangle-x^{3}_{0}, 
\end{array}
\end{eqnarray} 
%%%%%%%%%%%%%%%%%%%%
where  from definition (\ref{holomo.})
we have $x=\frac{1}{\sqrt{2\omega M_{\textsc p}}}(A+\bar A)$.  Hence, with
an eye on the  definition of $x_{0}$ in (\ref{a,n}), we obtain
  
%%%%%%%%%%%%%%%%%%%%
\begin{eqnarray}\label{qub.scal.quantiz} 
\left\langle a^{3} \right\rangle= \frac{3M(n+\frac{1}{2})}{M^{2}_{\textsc
p}\omega^{2}}+ \left(\frac{M}{M_{\textsc p} \omega}\right)^{3}.
 \end{eqnarray}
 %%%%%%%%%%%%%%%%%%%%
  Eqs.  (\ref{mass.quantize}) and (\ref{qub.scal.quantiz}) lead us to
%%%%%%%%%%%%%%%%%%%%
 \begin{eqnarray}
 \left\langle a^{3} \right\rangle=\frac{5\sqrt{2}}{({M_{\textsc p}\omega})^{\frac{3}{2}}} \left({n+\frac{1}{2}}\right)^\frac{3}{2}.
 \end{eqnarray}
 %%%%%%%%%%%%%%%%%%%%
Therefore,  the phase space average of squared  apparent horizon becomes $\left\langle R^{2}_{
\text{ah}}\right\rangle=\left(\frac{30\pi^{2}}{\omega^{2}}\right)\left(n+\frac{1}{2}\right)$, which shows that the area of the apparent horizon is quantized 
%%%%%%%%%%%%%%%%%%%%
\begin{eqnarray}\label{apperant dust}
  \left\langle\mathcal A_{\text {ah}}\right\rangle=4\pi\left\langle R^{2}_{
\text{ah}}\right\rangle= \left(\frac{5L^2_\textsc{p}}{6\pi
  }\right) \left(n+\frac{1}{2}\right).
  \end{eqnarray}
  %%%%%%%%%%%%%%%%%%%%
  Furthermore, from Eqs. (\ref{mass.quantize}) and (\ref{apperant dust})
  the total mass of the universe is  
%%%%%%%%%%%%%%%%%%%%
 \begin{eqnarray}\label{mass dust}
 {M}=M_{\textsc p}\frac{12\pi}{\sqrt{10}}\left(\frac{\langle\mathcal A_{\text{ah}}\rangle}{4G}\right)^{\frac{1}{2}}.
 \end{eqnarray}
  %%%%%%%%%%%%%%%%%%%%
 Substituting (\ref{mass dust}) into (\ref{entropy dust}),  the entropy
of dust will be
%%%%%%%%%%%%%%%%%%%%
 \begin{eqnarray}\label{entropy dust final}
 S^{(\text{dust})}\simeq \frac{M_{\textsc p}}{m}\left(\frac{\langle
\mathcal A_{\text{ah}}\rangle}{4G}\right)^{\frac{1}{2}}.
 \end{eqnarray}
 For further simplification, we use the well known relation between the radius 
 of universe (herein the radius  of apparent horizon defined via $L_{\text{ah}}:=\sqrt{\langle\mathcal A_{\text{ah}}\rangle}/4\pi$) and mass of nucleons, $m$, as a result of the
uncertainty
principle 
 \cite{Sivaram}
 \begin{eqnarray}\label{p1}
 L_{\text{ah}}\simeq\sqrt{N}\frac{1}{m}=\sqrt{\frac{M}{m}}\frac{1}{m}.
 \end{eqnarray}
 By substituting Eqs. (\ref{mass.quantize}) and (\ref{apperant dust}) in Eq. (\ref{p1}) we obtain
 \begin{eqnarray}\label{p2}
 m\simeq M_\text{P}(n+\frac{1}{2})^{-\frac{1}{6}}.
 \end{eqnarray}
 Also combining Eqs. (\ref{apperant dust}), (\ref{entropy dust final}) and (\ref{p2})
 we obtain
 \begin{eqnarray}
 S^{(\text{dust})}\simeq \left(\frac{\langle
 \mathcal A_{\text{ah}}\rangle}{4G}\right)^{\frac{2}{3}},
 \end{eqnarray}
 %%%%%%%%%%%%%%%%%%%%
 which again is in agreement with 't Hooft conjecture. Let us investigate this result for  large values of  quantum number $n$,
 which according to the correspondence principle,  the behavior
of the model should reduce to  its corresponding  classical region. For very large values of $n$, we can estimate from relation  (\ref{entropy quantize}) the following value for
the entropy of radiation%%%%%%%%%%%%%%%%%%%%
\begin{eqnarray}\label{compare entropy}
S_{\gamma}\simeq n^\frac{3}{4}.
\end{eqnarray}  
 %%%%%%%%%%%%%%%%%%%%
 On the other hand, the entropy of the dust content
of the universe will be
\begin{eqnarray}\label{lastt}
S^{(\text{dust})}\simeq n^\frac{2}{3}.
\end{eqnarray}
 Let us examine our
model for present epoch of  the universe. The current
entropy density of radiation in the universe
is $s_{0\gamma}=2970(\frac{T_{0}}{2.5K})^{3}\frac{1}{\text{cm}^{3}}
$. Therefore, the entropy of radiation  is $S_{0 \gamma}\simeq10^{88}$.
This estimation leads us to obtain the approximate value of the quantum number  $n$ as $n\simeq10^{117}$.
 Hence
by  inserting the obtained value of the quantum number $n$ in
Eq. (\ref{lastt}), we obtain $S^{(\text{dust})}\simeq 10^{79}$. 
This is in agreement with the classical estimation of the entropy of dust in the universe \cite{farmpton}. At the end of this section, let us
concentrate on the relation of our simple quantum cosmology model with the
Large Number Hypothesis (LNH). For very large values of quantum number $n$, Eqs.
(\ref{mass.quantize}), (\ref{apperant dust}) and (\ref{p2}) simplify to the
following well known scaling relations
\begin{eqnarray}\label{p4}
\begin{cases}
M\simeq\beta^3M_\text{P},\\
L_\text{ah}\simeq\beta^3L_\text{P},\\
m\simeq\beta^{-1}M_\text{P},
\end{cases}
\end{eqnarray}
where $\beta:=n^{1/6}\simeq10^{19}$. As showed by Marugan and
Carneiro  \cite{LNH}, the scaling relations that lie behind the LNH can be expressed in the same way as the above relations. Also, they have shown that if one assumes a flat universe dominated by the cosmological constant $\Lambda$, then Dirac's LNH can be explained in terms of the holographic conjecture.
On the other hand, our results show that the CHP could be the result of quantum
nature of the universe. Consequently it seems to be natural that the LNH could
be embedded
in quantum cosmology as one can see in relations  (\ref{p4}). Eliminating
 $\beta$ from the two last scaling relations in (\ref{p4}), we obtain
\begin{eqnarray}\label{p5}
m\simeq\left(\frac{1}{GL_{\text{ah}}}\right)^\frac{1}{3},
\end{eqnarray}
This equation is equivalent to the empirical Weinberg
formula for the mass of the pion \cite{Weinberg}.
%%%%%%%%%%%%%%%%%%%%%%%%%%%%%%%%%%%%%%%%%%%%%%%%%%%%%%%%%%%%%%%%%%%%%%%%%%%%%%%%%%%  
  \section{Conclusion}
 In this paper we studied the deformation quantization or phase space quantization
 of a closed quantum FLRW model, whose matter is either a fluid of radiation or dust. Our results show that the  peaks of
the WF  coincides with the classical trajectory of the universe.   
  Our main upshot is that  the CHP can be achieved by
  means of quantization of cosmological models. According to the CHP the entropy of  non-black hole configurations is given by relation $S\simeq(\frac{\mathcal{A}}{4G})^{\frac{3}{4}}$, where ${\mathcal{A}}$ denotes the area of containing volume. We showed
that the same result is maintained for radiation dominated universe, where
${\mathcal A}$ is replaced by the phase space average of apparent horizon $\langle{\mathcal A}_{\text{ah}}\rangle$, and $S$ is the total entropy (inside and outside). On
the other hand, for a dust dominated universe, we obtained $S\simeq(\frac{\langle\mathcal{A_{\text{ah}}}\rangle}{4G})^{\frac{2}{3}}$.
It seems that the power of apparent horizon
in units of Planck's surface is different for various matter configurations: for black holes this value is equal to 1, for radiation it is equal to  3/4, and for dust it is equal to 2/3.
We are aware that our
results are obtained within a very simple cosmological model. Nevertheless,
we think they are intriguing and provide
motivation for subsequent research works. Possible extensions
to test the  CHP  may include 
\begin{itemize}
\item 
Considering various Bianchi cosmological models.
\item 
Considering other perfect fluids besides radiation
and dust.
\item
Exploring the modified theories of gravity, like string cosmology and $f(R)$
theories.
\end{itemize}

\section{Acknowledgments} The authors would like to thank the anonymous
referee for  enlightening comments.
 %%%%%%%%%%%%%%%%%%%%%%%%%%%%%%%%%%%%%%%%%%%%%%%%%%%%%%%%%%%%%%%%%%%%%%%%%%%%%%%%%%%%% 

%+Bibliography

%-Bibliography

%-Bibliography

\end{document}